\documentclass{article}

\usepackage[utf8]{inputenc}
\usepackage[T1]{fontenc}

\usepackage{arxiv}          

\usepackage{amsmath}
\usepackage{amssymb}
\usepackage{amsfonts}

\usepackage{graphicx}
\usepackage{booktabs}
\usepackage{array}
\usepackage{xcolor}

\usepackage[numbers,sort&compress]{natbib}

\usepackage{hyperref}
\usepackage{orcidlink}

\definecolor{linknavy}{RGB}{20,60,140}
\hypersetup{%
    colorlinks=true,
    linkcolor=linknavy,
    citecolor=linknavy,
    urlcolor=linknavy,
    pdfauthor={Shrenik Jadhav, Zheng Liu},
    pdftitle={Digital Twin-Based Cooling System Optimization for Data Center},
    pdfkeywords={Cooling system optimization, data center, digital twin, supercomputer, energy efficiency},
}

\renewcommand{\headeright}{A Preprint}
\renewcommand{\undertitle}{A Preprint}
\renewcommand{\shorttitle}{Digital Twin-Based Cooling Optimization for Data Centers}

\begin{document}

\title{Digital Twin-Based Cooling System Optimization for Data Center}

\author{%
  Shrenik Jadhav\,\orcidlink{0009-0003-6906-7465} \\
  Department of Computer and Information Science \\
  University of Michigan-Dearborn \\
  Dearborn, USA \\
  \And
  Zheng Liu\,\orcidlink{0000-0003-4869-8893}\thanks{Corresponding author.} \\
  Department of Industrial and Manufacturing Systems Engineering \\
  University of Michigan-Dearborn \\
  Dearborn, USA \\
  \texttt{zhengtl@umich.edu} \\
}

\date{\today}

\maketitle

\begin{abstract}
Data center cooling systems consume significant auxiliary energy, yet optimization studies rarely quantify the gap between theoretically optimal and operationally deployable control strategies. This paper develops a digital twin of the liquid cooling infrastructure at the Frontier exascale supercomputer, in which a hot-temperature water system comprises three parallel subloops, each serving dedicated coolant distribution unit clusters through plate heat exchangers and variable-speed pumps. The surrogate model is built based on Modelica and validated through one full calendar year of 10-minute operational data following ASHRAE Guideline. The model achieves a subloop coefficient of variation of the root mean square error below 2.7\% and a normalized mean bias error within $\pm$2.5\%. Using this validated surrogate model, a layered optimization framework evaluates three progressively constrained strategies: an analytical flow-only optimization achieves 20.4\% total energy saving, unconstrained joint optimization of flow rate and supply temperature demonstrates 30.1\% total energy saving, and ramp-constrained optimization of flow rate and supply temperature, enforcing actuator rate limits, can reach total energy saving of 27.8\%. The analysis reveals that the baseline system operates at 2.9 times the minimum thermally safe flow rate, and the co-optimizing supply temperature with flow rate nearly doubles the savings achievable by flow reduction alone.
\end{abstract}


\section{Introduction} \label{s:intro}

Global data center electricity consumption reached 415~TWh in 2024, approximately 1.5\% of worldwide electricity demand, and is projected to exceed 945~TWh by 2030 \citep{IEA2025}. In the United States alone, data centers consumed 176~TWh in 2023 (4.4\% of national electricity), with projections ranging from 325 to 580~TWh by 2028 \citep{Shehabi2024}. While advances in server efficiency and virtualization have partially offset demand growth \citep{Masanet2020}, the proliferation of artificial intelligence workloads and the push toward exascale computing are placing unprecedented strain on cooling infrastructure. Cooling systems account for 30 - 40\% of total data center electricity consumption \citep{Zhang2021, Ebrahimi2014}, making them the single largest controllable load in most facilities. Yet the global average Power Usage Effectiveness (PUE) has stagnated between 1.55 and 1.59 since 2020 \citep{UptimeInstitute2024}, suggesting that fundamentally new optimization approaches are needed. While data center infrastructure efficiency, carbon usage effectiveness, supply heat index, rack cooling indicator, and water usage effectiveness are also becoming the important parts of evaluation matrix \citep{petrongolo2020simulation, zhu2023novel}.

High-performance computing (HPC) facilities intensify this challenge. Exascale supercomputers such as Frontier at Oak Ridge National Laboratory operate at power levels of 8 - 30~MW, employ 100\% direct liquid cooling with variable-speed pumps, and experience rapid thermal transients driven by workload dynamics \citep{Sun2024, Karimi2024}. These systems differ structurally from enterprise data centers: cooling loops have response times measured in minutes, and equipment constraints such as pump ramp-rate limits are operationally binding. Despite the scale of energy involved, systematic cooling system optimization for HPC facilities remains largely unexplored. Existing work on Frontier has focused on operational data collection \citep{Sun2024}, power management \citep{Karimi2024}, thermal stability \citep{Grant2026}, waste heat recovery \citep{Wang2024P}, and data-driven cooling inefficiency diagnosis \citep{Jadhav2026}. However, no study has coupled a physics-based digital twin of an exascale cooling system with systematic optimization, nor has any work formally quantified how much of the theoretical energy savings survive when practical operational constraints are imposed.

This paper addresses that gap through three contributions: (i)~a validated Modelica-based digital twin of Frontier's liquid cooling system, (ii)~a layered optimization framework that progressively reveals the sources and magnitudes of energy savings, and (iii)~a formally defined implementability gap metric that quantifies the energy savings lost when theoretical optima must accommodate practical constraints.

\begin{figure*}[t]
	\centering
	\includegraphics[width=0.85\textwidth]{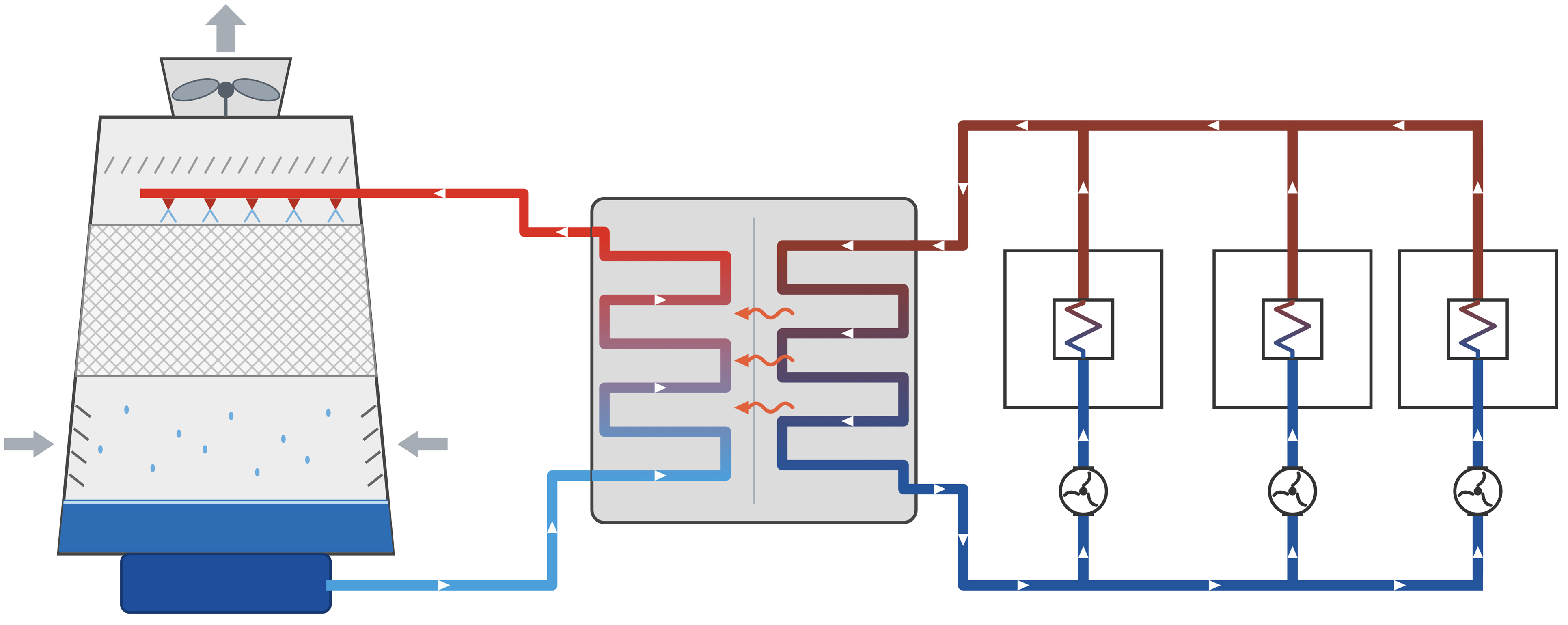}
\caption{Schematic of the Frontier supercomputer cooling architecture. The hot-temperature water system comprises three parallel subloops, each serving dedicated CDU clusters through plate heat exchangers. Each active subloop is driven by an independent 350-HP variable-speed pump. Warm return flow is rejected to the atmosphere via mechanical-draft cooling towers through an aggregate economizer heat exchanger. State variables used in the digital twin model are annotated at measurement points.}\label{fig:cooling_schematic}
\end{figure*}

\subsection{Data center cooling system optimization: approaches and limitations} \label{s:intro.optimization}

Research on data center cooling system optimization has accelerated considerably since a 40\% reduction in Google's cooling energy was reported using neural network recommendations \citep{Evans2016}, though that claim was subsequently critiqued for its lack of peer review and reproducibility \citep{Lazic2018}. The resulting body of work spans three broad methodological families. Reinforcement learning (RL) approaches treat the cooling system as a Markov decision process: deep RL has shown promising results on simulated workloads \citep{Li2020}, multi-agent RL integrated with a digital twin has demonstrated coordinated cooling control \citep{Sarkar2024}, and 14 - 21\% energy savings have been reported from offline RL deployed in a production data center \citep{Zhan2025}. However, RL methods face challenges around sample efficiency, safety during exploration, and policy interpretability that are particularly acute in safety-critical HPC environments. Model predictive control (MPC) formulates cooling optimization as a receding-horizon constrained problem; a rigorously documented 17.9\% cost reduction was achieved at Google's data centers \citep{Lazic2018}, and MPC has been applied to building climate control \citep{Oldewurtel2012} and chiller plant operation \citep{Ma2012}. MPC naturally handles constraints but degrades substantially with model uncertainty, with rule-based control shown to outperform MPC when uncertainty exceeds approximately 67\% \citep{Maasoumy2014}. Researchers also developed neural-network-based surrogate to predict server inlet temperatures in data center and couples it with a cooling system model to enable real-time optimization \citep{de2012neural}.
Classical gradient-based optimization, the workhorse of engineering design \citep{Kraft1988, gugulothu2026integrating}, is deterministic, computationally transparent, and well suited to cooling systems governed by smooth physics, yet it is notably absent from the data center cooling literature. Data-driven surrogates have also been applied to diagnose cooling inefficiency at HPC facilities \citep{Jadhav2026}, but these approaches identify waste rather than prescribing constrained optimal control.

Table~\ref{tab:lit_comparison} in Appendix~A summarizes the positioning of representative works. No existing paper simultaneously addresses physics-based digital twin modeling, systematic optimization, HPC liquid cooling, explicit practical constraints, and a formal implementability metric.

\subsection{Digital twins for thermal system modeling and optimization} \label{s:intro.digitaltwin}

The digital twin concept, originally articulated for product lifecycle management \citep{Grieves2017} and formalized for industrial applications \citep{Tao2019}, has found growing adoption in data center energy systems. The IISE community has recognized digital twins as a transformative methodology for design and manufacturing \citep{Knapp2020, Tan2024}. For optimization applications where control strategies must be explored beyond the range of historical operating data, physics-based digital twins offer superior generalization and interpretability compared to purely data-driven alternatives \citep{Afram2014}.

The Modelica language and its associated Buildings Library \citep{Wetter2014} have emerged as the dominant open-source platform for equation-based modeling of HVAC and cooling systems. Modelica-based data center cooling models have been developed for air-cooled facilities \citep{Fu2019a, Fu2019b}, district cooling systems \citep{Hinkelman2022}, chiller plants with water-side economizers \citep{Fan2021}, and comparative co-design of cooling controls \citep{Grahovac2023}. More recently, the Buildings Library has been coupled with EnergyPlus through the Spawn framework \citep{Wetter2024}.

For HPC cooling specifically, the Exascale Digital Twin (ExaDigiT) project at Oak Ridge National Laboratory developed a digital twin framework for liquid-cooled supercomputers using the TRANSFORM library within the Modelica environment \citep{Brewer2024, Kumar2024, Greenwood2024}. However, ExaDigiT focused on verification and validation rather than systematic optimization, a pattern that recurs across the literature: several groups have built physics-based cooling models or digital twin frameworks \citep{Fu2019a, Hinkelman2022, Athavale2024, Sarkar2024}, but the step from validated model to actionable optimization is frequently omitted. The present work constructs an independent Modelica-based digital twin using the Buildings Library \citep{Wetter2014}, purpose-built for steady-state optimization. The model is calibrated against 47{,}186 validated operational records \citep{Sun2024} and validated to ASHRAE Guideline~14 \citep{ASHRAE2014} criteria, achieving CV-RMSE of 1.96 - 2.67\% and NMBE within $\pm$2.5\% across all subloops (Section~\ref{s:methods.validation}).

\subsection{The implementability gap: from theoretical optima to practical control} \label{s:intro.implgap}

A persistent challenge in cooling system optimization is the gap between what is theoretically optimal and what can be practically implemented. This gap arises from actuator limitations, thermal inertia, safety margins, and operational protocols requiring gradual system behavior. Despite the practical importance of these constraints, the literature lacks a formal framework for quantifying their effect on achievable energy savings.

Related concepts exist in adjacent domains but do not address cooling system control. The building energy performance gap documents 20 - 550\% discrepancies between design-stage predictions and as-built operation \citep{deWilde2014, VanDronkelaar2016}, but this concerns modeling fidelity rather than the cost of operational constraints. The simulation-to-real gap in robotics and RL describes controller degradation when transferring from simulation to physical systems \citep{Salvato2021, Zhao2020, Chen2020}, but without defining a quantitative metric for cooling applications. The closest quantitative analog comes from energy storage optimization, where ramp-rate constraints reduce arbitrage profit by 10 - 35\%, with constrained solutions retaining 65 - 90\% of unconstrained profit \citep{Hashmi2024}.

We address this gap by introducing the implementability gap, defined as the fraction of theoretical optimal energy savings lost due to practical constraints, and its complement the recovery ratio (Section~\ref{s:methods.opt.gap}). The framework is operationalized through a layered optimization: Strategy~A optimizes only pump flow to establish a conservative baseline; Strategy~B co-optimizes pump flow and supply temperature without operational constraints, revealing the theoretical maximum; and Strategy~C re-introduces ramp-rate limits on all actuators, yielding the practically implementable savings. The difference between Strategies~B and~C directly measures the implementability gap.

\begin{figure*}[t]
	\centering
	\includegraphics[width=\textwidth]{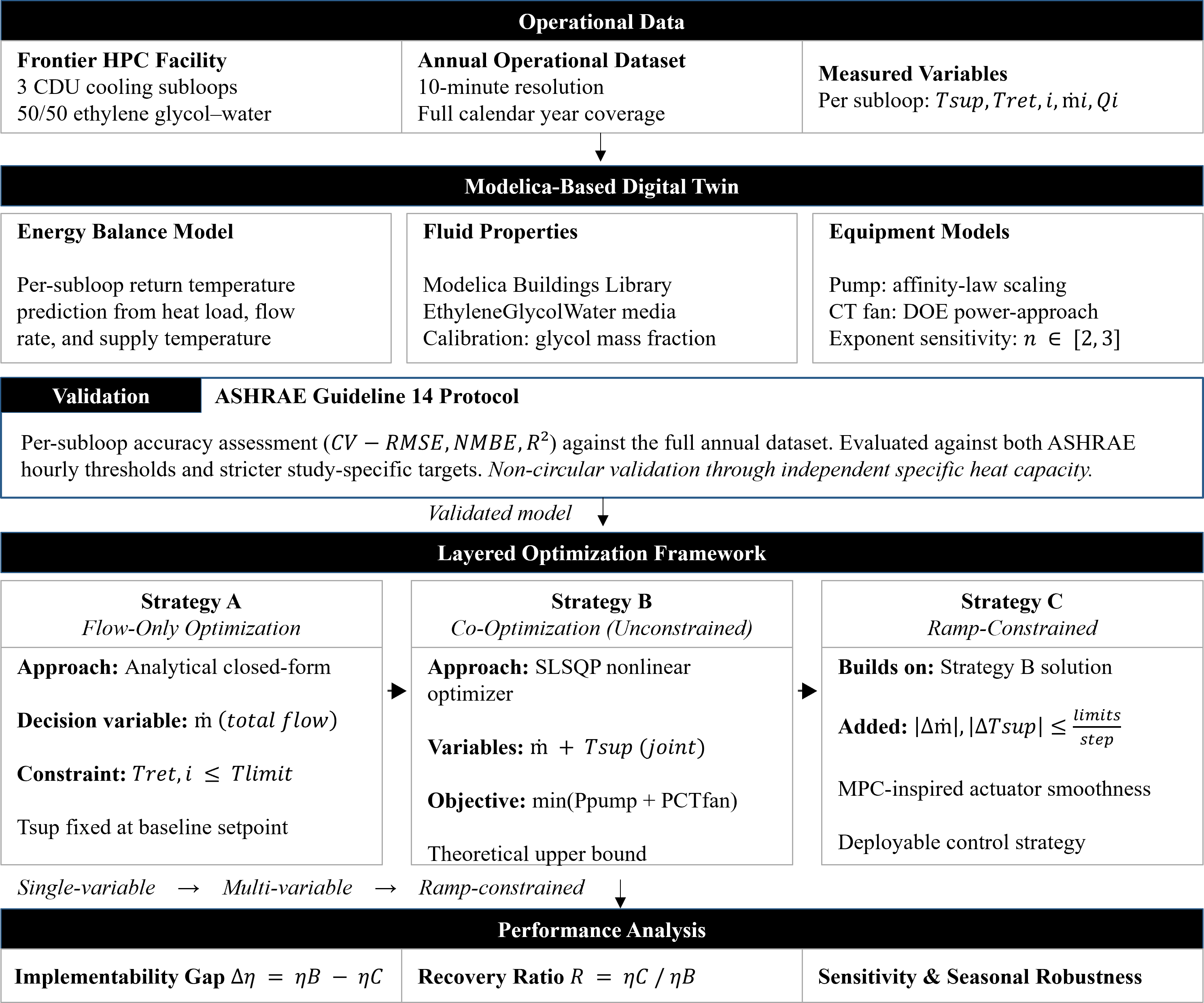}
	\caption{Digital twin framework for cooling system optimization. The physical layer provides operational data, the Modelica-based digital twin predicts thermal and energy performance, and the optimization layer determines energy-minimal control setpoints subject to thermal safety constraints.}
	\label{fig:framework}
\end{figure*}

\subsection{Contributions and paper organization} \label{s:intro.contributions}

The contributions of this paper are as follows:
\begin{enumerate}
	\item A validated Modelica-based digital twin of the Frontier supercomputer's three-subloop liquid cooling system, calibrated against one year of operational data (47{,}186 records) with CV-RMSE of 1.96 - 2.67\%.

	\item A layered optimization framework (flow-only, unconstrained co-optimization, and ramp-constrained co-optimization) using SLSQP, representing the first application of classical gradient-based optimization to HPC cooling systems.

	\item The implementability gap metric and recovery ratio, formal measures of the energy savings lost and retained when ramp-rate constraints are imposed on theoretically optimal strategies.

	\item Operational insights including 20.4\% savings from flow-only optimization, 30.1\% from unconstrained co-optimization, and 27.8\% from ramp-constrained co-optimization (92.4\% recovery ratio), demonstrating that the majority of theoretical savings are practically recoverable.
\end{enumerate}

Section~\ref{s:methods} presents the digital twin model and optimization formulation. Section~\ref{s:results} presents validation and optimization results. Section~\ref{s:discussion} discusses insights, robustness, and limitations. Section~\ref{s:Conclusion} concludes.

\section{Methodology} \label{s:methods}

This section presents the digital twin modeling framework and optimization formulation developed for the Frontier supercomputer cooling system. Section~\ref{s:methods.system} describes the physical cooling architecture and the digital twin concept. Section~\ref{s:methods.physics} formulates the physics-based sub-models. Section~\ref{s:methods.validation} defines the calibration procedure and validation criteria. Section~\ref{s:methods.optimization} presents the three optimization strategies in a layered formulation.

\subsection{System description and digital twin architecture} \label{s:methods.system}

\subsubsection{Frontier cooling system} \label{s:methods.system.frontier}

The Frontier supercomputer at Oak Ridge National Laboratory (ORNL) is an exascale high-performance computing (HPC) system, achieving 1.102~exaFLOPS on the HPL benchmark \citep{Sun2024}. Frontier employs a warm-water liquid cooling architecture comprising three thermally coupled fluid loops arranged in a cascaded heat rejection path, with operational waste heat ranging from 8 to 28~MW depending on computational workload, as illustrated in Figure~\ref{fig:cooling_schematic}.

The tertiary loop (also called the coolant distribution unit, or CDU, loop) circulates a 50/50 ethylene glycol-water mixture (reported as a 50\% solution with the basis unspecified \citep{Sun2024}; see Section~\ref{s:methods.validation.calibration} for calibration details) directly through the compute cabinets. Twenty-five CDUs serve 74 compute cabinets, rejecting waste heat from the processors and accelerators into the glycol coolant. The heated glycol is then passed through plate heat exchangers that transfer the thermal energy to the secondary loop (the hot-temperature water (HTW) loop), which circulates pure water at flow rates between 113 and 412~kg/s depending on operating conditions. The HTW loop in turn rejects heat to the primary loop (the cooling tower water (CTW) loop) through a second set of heat exchangers, and the CTW loop dissipates the heat to the atmosphere via mechanical-draft cooling towers.

A defining feature of the Frontier cooling architecture is that the HTW system is organized into four parallel subloops, of which three are active and one is held in reserve \citep{Sun2024}. Each active subloop is driven by an independent 350-HP variable-speed centrifugal pump equipped with a variable-frequency drive (VFD), for a total of four installed pumps (three active, one standby). Because each pump has its own power-flow characteristic and operates on a separate VFD, the three active pumps are modeled independently in the digital twin (Section~\ref{s:methods.validation.modelica}). Each active subloop contains its own CDU cluster, dedicated pump, and heat exchanger. The nominal flow distribution across the three active subloops is approximately 24.6\%, 26.0\%, and 49.5\% of the total HTW flow, as derived from the operational dataset and verified by the Modelica simulation (Section~\ref{s:results}). These flow fractions are determined by the hydraulic resistance of each subloop's piping network and remain approximately constant across the full operating range; they are fixed design-time parameters governed by pipe geometry rather than actively controlled variables. This parallel topology is critical to the optimization formulation because the thermal constraint must be satisfied independently in each branch: the return temperature in every subloop must remain below the equipment limit regardless of conditions in the other two branches. The fourth (reserve) subloop is maintained at standby pressure but carries negligible flow during normal operation; its parasitic contribution is excluded from the energy accounting.

The operational dataset provides records at 10-minute intervals \citep{Sun2024} spanning the full calendar year 2023. The dataset includes supply and return temperatures, per-subloop flow rates and heat loads, and aggregate power consumption. After removing sensor anomalies, maintenance shutdowns, and physically inconsistent records (return temperature below supply temperature, heat load exceeding 15~MW, supply temperature below 5$^\circ$C, or total flow below 30~kg/s), 47{,}186 validated operating points remain for model development and optimization.

\subsubsection{Digital twin framework} \label{s:methods.system.twin}

The digital twin developed in this work serves as a simulation-based optimization testbed: a calibrated physics model that can evaluate candidate control strategies offline, using historical operating data as boundary conditions, before any physical implementation. Figure~\ref{fig:framework} illustrates the overall framework, which consists of three layers.

The physical layer comprises the Frontier cooling plant and its supervisory control system, which currently operates the HTW pumps at fixed-speed setpoints. The digital twin layer is a physics-based Modelica model implemented in OpenModelica using components from the Buildings Library \citep{Wetter2014}. The model ingests measured boundary conditions (heat loads, supply temperature, and flow rates) and predicts system-level thermal and energy performance. The optimization layer formulates and solves energy minimization problems using the digital twin's analytical prediction equations as the plant model, subject to thermal safety constraints derived from equipment specifications.

The key advantage of this architecture is that the optimization operates on the same physics equations used in the validated digital twin, ensuring consistency between the model used for control design and the model used for performance prediction. This avoids the simulation-to-real gap that arises when surrogate models (e.g., neural networks) are used for optimization but validated against a different simulator.

\subsection{Physics-based model formulation} \label{s:methods.physics}
The digital twin represents the Frontier HTW cooling system using three coupled sub-models: heat exchangers connecting the CDU, HTW, and CTW loops; variable-speed pumps driving the HTW flow; and mechanical-draft cooling towers rejecting heat to the atmosphere. Each sub-model is formulated below, with notation defined at first use.

\subsubsection{Heat exchanger model} \label{s:methods.physics.hx}

Each subloop's CDU-to-HTW heat exchanger is modeled using the effectiveness-NTU ($\varepsilon$-NTU) method \citep{Incropera2007}. For a counterflow plate heat exchanger with known effectiveness $\varepsilon_i$, the return temperature in subloop $i$ at timestep $t$ is determined by the energy balance:
\begin{equation} \label{eq:energy_balance}
	T_{\mathrm{ret},i}(t) = T_{\mathrm{sup}}(t) + \frac{Q_i(t)}{f_i \cdot \dot{m}(t) \cdot c_p}
\end{equation}
where $f_i$ is the fraction of total flow through subloop $i$. This expression follows directly from the first law of thermodynamics applied to the coolant stream passing through the CDU cluster, under the assumption that all waste heat $Q_i$ is absorbed by the HTW fluid. The flow fractions are determined by the parallel hydraulic network topology and remain approximately constant at $f_1 = 0.244$, $f_2 = 0.258$, and $f_3 = 0.498$ across the full operating range, as derived from the operational dataset and confirmed by the Modelica hydraulic model (Section~\ref{s:results}).

The HTW-to-CTW heat exchangers connecting the secondary and primary loops are modeled using a constant-effectiveness formulation (\texttt{ConstantEffectiveness} component in the Modelica Buildings Library; \citep{Wetter2014}), which represents the plate heat exchangers as idealized devices with prescribed thermal effectiveness:
\begin{equation} \label{eq:epsilon_ntu}
	Q_{\mathrm{HX}} = \varepsilon \cdot C_{\min} \cdot (T_{\mathrm{hot,in}} - T_{\mathrm{cold,in}})
\end{equation}
where $C_{\min} = \min(\dot{m}_{\mathrm{hot}} c_{p,\mathrm{hot}},\; \dot{m}_{\mathrm{cold}} c_{p,\mathrm{cold}})$ is the minimum heat capacity rate and $\varepsilon$ is the heat exchanger effectiveness. The nominal effectiveness values used in the model are $\varepsilon_{\mathrm{CDU\text{-}HTW}} = 0.65$ and $\varepsilon_{\mathrm{HTW\text{-}CTW}} = 0.75$, consistent with the temperature profiles observed in the Frontier dataset.

To enable closed-form optimization (Section~\ref{s:methods.optimization}), the heat exchanger thermal resistance is characterized by an aggregate overall heat transfer coefficient-area product $UA$. Under the assumption that the HTW-side convective resistance dominates the overall thermal resistance (i.e., wall conduction and CTW-side resistance are comparatively small), the Dittus-Boelter correlation for turbulent internal flow yields $h \propto \dot{m}^{0.8}$ \citep{Incropera2007}, and consequently:
\begin{equation} \label{eq:ua_scaling}
	UA(\dot{m}) = UA_{\mathrm{nom}} \left( \frac{\dot{m}}{\dot{m}_{\mathrm{nom}}} \right)^{0.8}
\end{equation}
where $UA_{\mathrm{nom}} = 2{,}252$~kW/K is the nominal value at the reference flow rate $\dot{m}_{\mathrm{nom}} = 190$~kg/s. This scaling relationship is used in the back-calculation of cooling tower water supply temperature from measured HTW conditions (Section~\ref{s:methods.validation}).

\subsubsection{Pump power model} \label{s:methods.physics.pump}

The HTW pumps are modeled using the affinity laws for variable-speed centrifugal pumps. For a closed-loop system with negligible static head (as in the Frontier HTW loop, where the pressure drop is dominated by pipe friction and component losses), the system curve is purely quadratic and the pump power relates to flow rate as:
\begin{equation} \label{eq:pump_power}
	P_{\mathrm{pump}}(\dot{m}) = P_{\mathrm{pump,nom}} \left( \frac{\dot{m}}{\dot{m}_{\mathrm{nom}}} \right)^n
\end{equation}

The baseline pump power exponent is $n = 3$, corresponding to the theoretical cubic law ($P \propto N^3$, $Q \propto N$, hence $P \propto Q^3$) for variable-speed operation against a purely quadratic system curve \citep{ASHRAE2020}. In practice, the exponent may deviate from 3 due to residual static head, valve characteristics, and motor efficiency variations. The sensitivity of the optimization results to the pump exponent is examined in Section~\ref{s:results} for $n \in \{2.0, 2.5, 3.0\}$.

The nominal pump power $P_{\mathrm{pump,nom}}$ is calibrated from the Frontier dataset by matching the measured total accessory power at the reference operating point. At the Frontier facility, the HTW pumps currently operate at fixed-speed setpoints: approximately 200~kg/s during the winter months (November through April) and 350 - 400~kg/s during the summer months (May through October). This step-like operating pattern, rather than a continuous modulation proportional to thermal demand, represents the primary source of energy waste that the proposed optimization addresses.
In the updated digital twin, each of the three active subloop pumps is modeled independently, with per-pump nominal power scaled by the subloop flow fraction. The total pump power is the sum across all three active pumps: $P_{\mathrm{pump}} = \sum_{k=1}^{3} P_{\mathrm{pump},k}(\dot{m}_k)$, where $\dot{m}_k = f_k \cdot \dot{m}$ is the flow through subloop $k$.

\subsubsection{Cooling tower model} \label{s:methods.physics.ct}

The mechanical-draft cooling towers reject heat from the CTW loop to the atmosphere. The cooling tower outlet temperature is modeled using a fixed-approach formulation:
\begin{equation} \label{eq:ct_outlet}
	T_{\mathrm{CT,out}} = T_{\mathrm{wb}} + \Delta T_{\mathrm{app}}
\end{equation}
where $T_{\mathrm{wb}}$ is the ambient wet-bulb temperature and $\Delta T_{\mathrm{app}}$ is the approach temperature, which is the minimum temperature difference between the cooled water leaving the tower and the ambient wet-bulb. A nominal approach of 4$^\circ$C is used, consistent with values reported for large-scale data center cooling installations \citep{ASHRAE2020}.

The cooling tower fan power is modeled as a function of the heat rejection rate and the temperature driving potential:
\begin{equation} \label{eq:ct_fan_power}
	P_{\mathrm{CT}} = P_{\mathrm{CT,nom}} \left( \frac{Q_{\mathrm{rej}}}{Q_{\mathrm{rej,nom}}} \right) \left( \frac{\Delta T_{\mathrm{app,nom}}}{\Delta T_{\mathrm{app}}} \right)^\gamma
\end{equation}
where $Q_{\mathrm{rej}}$ is the actual heat rejection rate, $Q_{\mathrm{rej,nom}}$ is the design heat rejection rate, and $\gamma = 1.0$ is the fan power scaling exponent. The heat rejected by the cooling tower equals the total IT heat load plus the pump work added to the fluid:
\begin{equation} \label{eq:heat_rejection}
	Q_{\mathrm{rej}} = \sum_{i=1}^{3} Q_i + P_{\mathrm{pump}}
\end{equation}

The total cooling system power consumption is the sum of pump and cooling tower fan power:
\begin{equation} \label{eq:total_power}
	P_{\mathrm{total}} = P_{\mathrm{pump}} + P_{\mathrm{CT}}
\end{equation}

This formulation captures a fundamental trade-off: reducing pump flow lowers $P_{\mathrm{pump}}$ (cubic relationship) but raises the return temperature, which narrows the approach temperature and increases $P_{\mathrm{CT}}$. The optimization must navigate this coupling to minimize total system energy.

\subsection{Model calibration and validation approach} \label{s:methods.validation}

\subsubsection{Modelica implementation} \label{s:methods.validation.modelica}

The physics-based model is implemented in OpenModelica 1.23 using the Modelica Buildings Library v11 \citep{Wetter2014}. The model represents the aggregate HTW-to-CTW economizer heat exchanger using the \texttt{PlateHeatExchangerEffectivenessNTU} component in counterflow configuration, which internally computes the heat exchanger effectiveness from the nominal operating point using the $\varepsilon$-NTU method and scales the overall heat transfer coefficient $UA$ with flow rate via the Dittus-Boelter exponent ($n = 0.8$). The nominal parameters used in the Modelica model are summarized in Table~\ref{tab:ehx_params}.
 
The three HTW pumps are modeled as independent flow-controlled movers (\texttt{FlowControlled\_m\_flow}), one per active subloop, each prescribing its measured per-subloop flow rate as a boundary condition. This configuration matches the physical system, in which each of the three active 350-HP pumps operates on an independent VFD with its own power-flow characteristic. Pump heat addition to the fluid is neglected (\texttt{addPowerToMedium = false}) because the pump work ($\sim$17~kW per pump) is three orders of magnitude smaller than the IT heat load ($\sim$9~MW).
 
The implementation further uses \texttt{Buildings.\allowbreak Fluid.\allowbreak FixedResistances.\allowbreak PressureDrop} for the hydraulic resistance elements that produce the target flow distribution across the three parallel branches, and \texttt{Buildings.\allowbreak Media.\allowbreak Antifreeze.\allowbreak EthyleneGlycolWater} for the glycol-water fluid properties based on polynomial correlations for secondary working fluids \citep{Melinder2010}. The HTW loop operates at a system gauge pressure of approximately 6.2~bar (620~kPa). An expansion vessel provides the pressure reference point for the closed-loop hydraulic circuit.

\begin{table}[htbp]\footnotesize\setlength{\tabcolsep}{4pt}
	\centering
	\caption{Economizer heat exchanger (EHX) parameters used in the Modelica digital twin. The aggregate EHX represents four physical plate heat exchangers operating in parallel.}
	\label{tab:ehx_params}
	\begin{tabular}{l l l}
		\toprule
		\textbf{Parameter} & \textbf{Value} & \textbf{Description} \\
		\midrule
		Configuration       & Counterflow      & Plate heat exchanger type \\
		$Q_{\mathrm{nom}}$  & 9{,}120 kW       & Nominal heat transfer rate \\
		$T_{a1,\mathrm{nom}}$ & 33.7$^\circ$C  & HTW inlet (return) temperature \\
		$T_{a2,\mathrm{nom}}$ & 17.5$^\circ$C  & CTW inlet (supply) temperature \\
		$\dot{m}_{1,\mathrm{nom}}$ & 190 kg/s  & HTW-side nominal flow rate \\
		$\dot{m}_{2,\mathrm{nom}}$ & 292 kg/s  & CTW-side nominal flow rate \\
		CTW/HTW flow ratio  & 1.535            & Mass flow ratio at nominal point \\
		$UA_{\mathrm{nom}}$ & 2{,}252 kW/K     & Nominal overall conductance \\
		$UA$ scaling exponent & 0.8            & Dittus-Boelter flow exponent \\
		$\Delta p_{\mathrm{HTW}}$ & 50 kPa     & HTW-side pressure drop \\
		$\Delta p_{\mathrm{CTW}}$ & 50 kPa     & CTW-side pressure drop \\
		$\Delta p_{\mathrm{branch}}$ & 80 kPa  & Per-subloop branch resistance \\
		$\Delta p_{\mathrm{header}}$ & 85 kPa  & Supply/return header resistance \\
		\bottomrule
	\end{tabular}
\end{table}

\subsubsection{Fluid property calibration} \label{s:methods.validation.calibration}

A single calibration parameter governs the model's accuracy: the effective glycol mass fraction $X_a$ in the ethylene glycol-water mixture. The original dataset reports a 50/50 ethylene glycol-water coolant without specifying whether this refers to a mass or volumetric ratio \citep{Sun2024}. coolant without specifying whether this refers to a mass or volumetric ratio. Analysis of the dataset reveals that the waste-heat columns are derived quantities, computed internally as $Q_i = \dot{m}_i \, c_p \, (T_{\mathrm{ret},i} - T_{\mathrm{sup}})$ with a fixed conversion constant of $c_p = 3{,}709$ J/(kg$\cdot$K) and a flow conversion factor of 0.0631 kg/(s$\cdot$gpm). This constant is applied uniformly across all 49{,}869 records with zero variance, confirming that $Q$ is not an independent measurement. Consequently, using $c_p = 3{,}709$ J/(kg$\cdot$K) to predict return temperatures from $Q$ would recover the dataset's own formula, a circular validation. 

To avoid this circularity, the digital twin adopts $c_p = 3{,}500$ J/(kg$\cdot$K), the standard engineering value for 50/50 ethylene glycol-water at operating temperatures (25 - 35$\,^{\circ}$C). In the Modelica model, the Buildings Library's \texttt{EthyleneGlycolWater} media package takes $X_a$ as a mass fraction and yields $c_p \approx 3{,}190$ J/(kg$\cdot$K) at $X_a = 0.50$ (reference temperature 30$\,^{\circ}$C). Setting $X_a = 0.40$ yields $c_p \approx 3{,}500$ J/(kg$\cdot$K), aligning the Modelica fluid properties with the adopted validation constant. This adjustment is physically consistent with an effective volumetric fraction of approximately 50\% when accounting for the density difference between ethylene glycol (1{,}113 kg/m$^3$) and water (998 kg/m$^3$) at 30$\,^{\circ}$C. The 6\% gap between the adopted $c_p$ and the dataset's internal constant (3{,}500 vs.\ 3{,}709) produces the validation errors reported in Section \ref{s:results.validation}; this gap represents a genuine test of the model's predictive capability rather than a numerical artefact. The sensitivity of the optimization results to $c_p$ uncertainty is minimal because the optimizer uses the same property correlations as the digital twin, so any systematic offset cancels in the relative comparison between baseline and optimized performance.

\begin{figure*}[t]
	\centering
	\includegraphics[width=\textwidth,height=0.42\textheight,keepaspectratio]{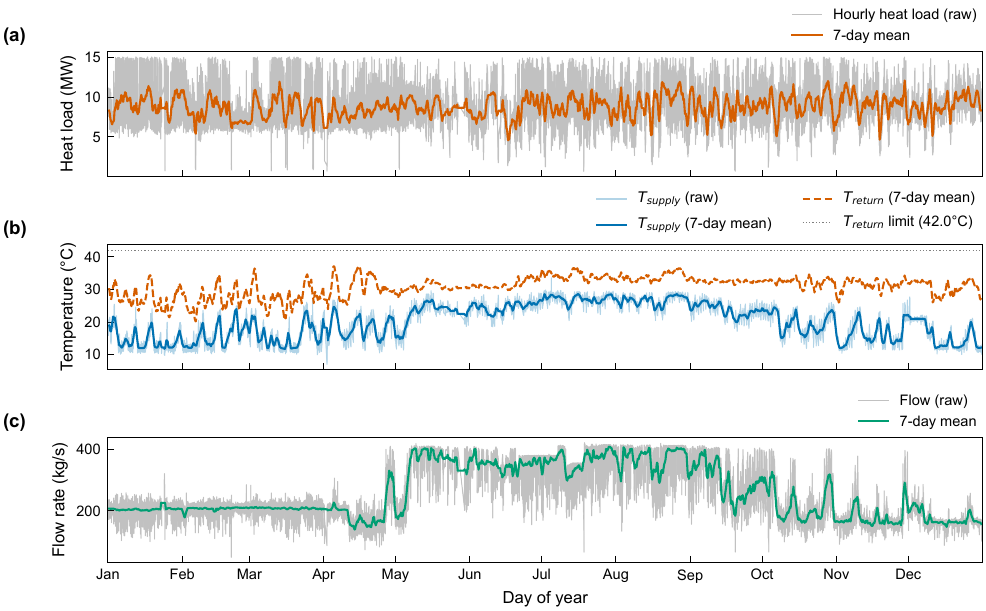}
	\caption{Annual operating conditions for the Frontier HTW cooling loop in 2023: (a) total computational heat load, (b) supply and return temperatures with 42$^\circ$C constraint threshold, and (c) total pump flow rate. Gray traces show 10-minute data; colored lines show 7-day rolling means.}
	\label{fig:annual_conditions}
\end{figure*}

\subsubsection{Validation protocol} \label{s:methods.validation.protocol}

The model is validated against measured data following ASHRAE Guideline 14-2014 \citep{ASHRAE2014}, which defines acceptance criteria for calibrated simulation models (note: superseded by Guideline 14-2023; the 2014 edition is used here for consistency with the majority of published digital twin validation studies). Two primary metrics are used:

The CV-RMSE quantifies the overall prediction scatter:
\begin{equation} \label{eq:cvrmse}
	\mathrm{CV\text{-}RMSE} = \frac{1}{\bar{y}} \sqrt{\frac{\sum_{t=1}^{n} \left( y_t - \hat{y}_t \right)^2}{n - p}} \times 100\%
\end{equation}

The Normalized Mean Bias Error (NMBE) quantifies systematic over- or under-prediction:
\begin{equation} \label{eq:nmbe}
	\mathrm{NMBE} = \frac{\sum_{t=1}^{n} (y_t - \hat{y}_t)}{(n-p) \cdot \bar{y}} \times 100\%
\end{equation}
where $y_t$ and $\hat{y}_t$ are the measured and predicted values at timestep $t$, $\bar{y}$ is the mean of the measured values, $n$ is the number of observations, and $p$ is the number of adjustable model parameters ($p = 1$ for calibrated simulation per ASHRAE Guideline 14 \citep{ASHRAE2014}). ASHRAE Guideline 14 specifies acceptance thresholds of CV-RMSE $\leq 30\%$ and $|\mathrm{NMBE}| \leq 10\%$ for hourly data. To ensure a stringent test of model fidelity, we adopt stricter targets of CV-RMSE $\leq 5\%$ and $|\mathrm{NMBE}| \leq 5\%$ for the per-subloop return temperature predictions.

The validation is conducted using a system identification approach: measured supply temperature $T_{\mathrm{sup}}(t)$, per-subloop flow rates $f_i \cdot \dot{m}(t)$, and per-subloop heat loads $Q_i(t)$ are prescribed as boundary conditions, and the model predicts the per-subloop return temperatures $T_{\mathrm{ret},i}(t)$. This isolates the heat transfer physics as the quantity being validated, removing uncertainty in pump performance and cooling tower behavior from the validation assessment. Two representative one-week periods are selected for detailed analysis: a summer week (July 1 - 7) representing peak thermal stress with supply temperatures reaching 33.6$^\circ$C and flows up to 412~kg/s, and a winter week (January 7th - 14th) representing part-load conditions with supply temperatures as low as 10.6$^\circ$C and flows down to 113~kg/s. Validation results are presented in Section~\ref{s:results.validation}.

\subsection{Optimization formulation} \label{s:methods.optimization}

The objective of the cooling system optimization is to minimize the total annual energy consumption of the HTW cooling infrastructure while maintaining thermal safety for the compute equipment. The equipment manufacturer specifies a maximum coolant return temperature of 45$^\circ$C. To provide an operational safety margin, the optimization enforces a constraint of $T_{\mathrm{ret},i} \leq 42^\circ$C in each subloop, corresponding to a 3$^\circ$C buffer below the equipment limit.

Three optimization strategies are formulated in a layered progression, where each subsequent strategy extends the preceding one by adding decision variables or constraints. This structure enables a systematic evaluation of the trade-offs between theoretical optimality, operational complexity, and implementability.

\subsubsection{Strategy~A: Flow-only optimization} \label{s:methods.opt.A}

Strategy~A optimizes only the HTW pump flow rate while keeping the supply temperature at its measured baseline value. The formulation is:
\begin{subequations} \label{eq:stratA}
\begin{align}
	\min_{m(t)} \quad & P_{\mathrm{total}}\bigl(m(t),\, T_{\mathrm{sup}}^{\mathrm{base}}(t)\bigr) \label{eq:stratA_obj} \\
	\text{s.t.} \quad & T_{\mathrm{ret},i}\bigl(m(t),\, T_{\mathrm{sup}}^{\mathrm{base}}(t)\bigr) \leq T_{\mathrm{limit}}, \quad i = 1,2,3 \label{eq:stratA_thermal} \\
	& m_{\min} \leq m(t) \leq m_{\max} \label{eq:stratA_bounds}
\end{align}
\end{subequations}
where $T_{\mathrm{limit}} = 42^\circ$C, $m_{\min} = 30$~kg/s (minimum stable pump operation), and $m_{\max} = 420$~kg/s (maximum pump capacity). The supply temperature $T_{\mathrm{sup}}^{\mathrm{base}}(t)$ is the measured value at each timestep and is not a decision variable in this formulation.

For Strategy~A, the thermal constraint \eqref{eq:stratA_thermal} admits an analytical solution. Substituting the energy balance \eqref{eq:energy_balance} into the constraint and solving for the minimum feasible flow yields:
\begin{equation} \label{eq:m_min_analytical}
	m_{\min}^{\mathrm{thermal}}(t) = \max_{i \in \{1,2,3\}} \frac{Q_i(t)}{f_i \cdot c_p \cdot \bigl(T_{\mathrm{limit}} - T_{\mathrm{sup}}^{\mathrm{base}}(t)\bigr)}
\end{equation}

\begin{figure*}[t]
	\centering
	\includegraphics[width=\textwidth]{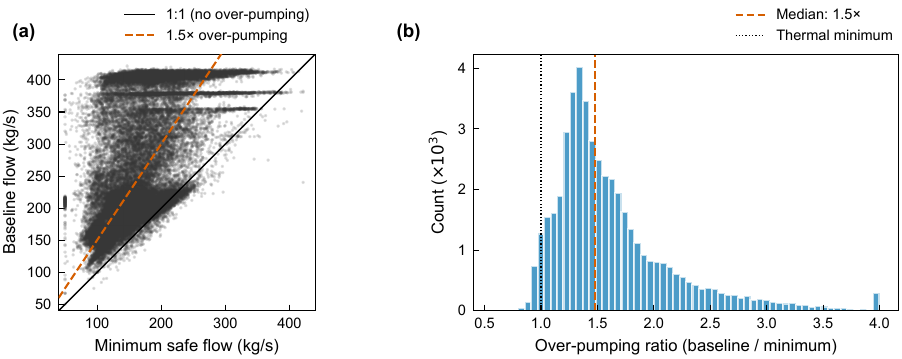}
	\caption{Baseline over-pumping diagnosis: (a) measured baseline flow versus analytically computed minimum safe flow, and (b) distribution of the over-pumping ratio. The median ratio of $1.5\times$ indicates systematic over-pumping across all operating conditions.}
	\label{fig:overpumping}
\end{figure*}

The binding constraint is determined by the subloop with the highest heat load relative to its flow fraction. In the Frontier system, the mean heat load fractions are $Q_1/Q_{\mathrm{total}} \approx 0.181$, $Q_2/Q_{\mathrm{total}} \approx 0.194$, and $Q_3/Q_{\mathrm{total}} \approx 0.625$, while the corresponding flow fractions are $f_1 = 0.246$, $f_2 = 0.260$, and $f_3 = 0.495$. Subloop~3 carries approximately 63\% of the total heat load through 49.5\% of the total flow, reflecting its higher mean temperature differential ($\overline{\Delta T}_3 = 13.8\,^{\circ}$C vs.\ $\overline{\Delta T}_{1,2} \approx 8\,^{\circ}$C). This makes it the binding branch at virtually every operating point. The optimal flow is then:
\begin{equation} \label{eq:stratA_solution}
	m^*(t) = \max\bigl\{ m_{\min},\; m_{\min}^{\mathrm{thermal}}(t) \bigr\}
\end{equation}

This closed-form solution eliminates the need for iterative numerical optimization and guarantees zero constraint violations by construction.

\subsubsection{Strategy~B: Unconstrained co-optimization} \label{s:methods.opt.B}

Strategy~B extends Strategy~A by introducing the supply temperature setpoint $T_{\mathrm{sp}}$ as a second decision variable. The physical mechanism is that raising the supply temperature reduces the heat rejection burden on the cooling towers (by narrowing the temperature difference that drives tower heat transfer), at the cost of reducing the thermal headroom available for pump flow reduction. The formulation is:
\begin{subequations} \label{eq:stratB}
\begin{align}
	\min_{m(t),\, T_{\mathrm{sp}}(t)} \quad & P_{\mathrm{total}}\bigl(m(t),\, T_{\mathrm{sp}}(t)\bigr) \label{eq:stratB_obj} \\
	\text{s.t.} \quad & T_{\mathrm{ret},i}\bigl(m(t),\, T_{\mathrm{sp}}(t)\bigr) \leq T_{\mathrm{limit}}, \quad i = 1,2,3 \label{eq:stratB_thermal} \\
	& m_{\min} \leq m(t) \leq m_{\max} \label{eq:stratB_flow} \\
	& T_{\mathrm{sp,min}} \leq T_{\mathrm{sp}}(t) \leq T_{\mathrm{sp,max}} \label{eq:stratB_temp}
\end{align}
\end{subequations}
where $T_{\mathrm{sp,min}} = 10^\circ$C and $T_{\mathrm{sp,max}} = 35^\circ$C bound the feasible supply temperature range. Unlike Strategy~A, the two-variable problem does not admit a simple closed-form solution because of the nonlinear coupling between $m$ and $T_{\mathrm{sp}}$ in both the pump power Equation (\ref{eq:pump_power}) and the cooling tower power Equations (\ref{eq:ct_fan_power} - \ref{eq:heat_rejection}). Strategy~B is solved using Sequential Least-Squares Programming (SLSQP) \citep{Kraft1988} at each timestep independently, treating each 10-minute interval as a quasi-steady-state optimization.

Strategy~B represents the theoretical upper bound on achievable savings because it optimizes both available control degrees of freedom without imposing actuator constraints. However, the resulting control trajectories may exhibit large step-to-step changes in both flow rate and supply temperature, which are impractical for real-time implementation due to water hammer risk, thermal shock to equipment, and variable-frequency drive slew rate limitations.

\subsubsection{Strategy~C: Ramp-constrained co-optimization} \label{s:methods.opt.C}

Strategy~C addresses the implementability gap by augmenting the Strategy~B formulation with ramp-rate constraints that limit the step-to-step change in both decision variables. The formulation is:
\begin{subequations} \label{eq:stratC}
\begin{align}
	\min_{m(t),\, T_{\mathrm{sp}}(t)} \quad & P_{\mathrm{total}}\bigl(m(t),\, T_{\mathrm{sp}}(t)\bigr) \label{eq:stratC_obj} \\
	\text{s.t.} \quad & T_{\mathrm{ret},i}\bigl(m(t),\, T_{\mathrm{sp}}(t)\bigr) \leq T_{\mathrm{limit}}, \quad i = 1,2,3 \label{eq:stratC_thermal} \\
	& m_{\min} \leq m(t) \leq m_{\max} \label{eq:stratC_flow} \\
	& T_{\mathrm{sp,min}} \leq T_{\mathrm{sp}}(t) \leq T_{\mathrm{sp,max}} \label{eq:stratC_temp} \\
	& \bigl| m(t) - m(t-1) \bigr| \leq \Delta m_{\max} \label{eq:stratC_ramp_m} \\
	& \bigl| T_{\mathrm{sp}}(t) - T_{\mathrm{sp}}(t-1) \bigr| \leq \Delta T_{\max} \label{eq:stratC_ramp_T}
\end{align}
\end{subequations}

The ramp-rate limits are set to $\Delta m_{\max} = 50$~kg/s per 10-minute step and $\Delta T_{\max} = 1^\circ$C per 10-minute step. These values are operational constraints reflecting thermal management requirements for large-scale hot-water distribution systems: the flow ramp limit of 50~kg/s per step corresponds to a maximum rate of change of 5~kg/s per minute, which provides adequate margin against water hammer and pressure transients in the piping network while remaining well within the response capability of modern variable-frequency drives \citep{ASHRAE2020}. The temperature ramp limit of 1$^\circ$C per step prevents thermal shock to the heat exchangers and ensures that downstream CDU control loops can track the setpoint change without oscillation.

The ramp constraints \eqref{eq:stratC_ramp_m} - \eqref{eq:stratC_ramp_T} introduce temporal coupling between successive timesteps: the feasible region at timestep $t$ depends on the solution at timestep $t-1$. This transforms the problem from a collection of independent per-timestep optimizations (as in Strategies~A and~B) into a sequential decision process. The formulation follows the principles of model predictive control (MPC), where the optimizer at each step accounts for the current state (including the previous setpoints) when computing the next control action. Strategy~C is solved sequentially using SLSQP with the ramp constraints reformulated as bound tightening: at each timestep, the feasible bounds on $m$ and $T_{\mathrm{sp}}$ are narrowed to $[m(t-1) - \Delta m_{\max},\; m(t-1) + \Delta m_{\max}]$ and $[T_{\mathrm{sp}}(t-1) - \Delta T_{\max},\; T_{\mathrm{sp}}(t-1) + \Delta T_{\max}]$, respectively, intersected with the global bounds.

A safety override is incorporated: if the thermal constraint is violated at the optimal solution (which can occur when the ramp limit prevents the flow from increasing fast enough after a sudden load spike), the flow rate is immediately increased to the minimum thermally safe value computed by Equation~\eqref{eq:m_min_analytical}, overriding the ramp constraint. This ensures that thermal safety always takes priority over actuator smoothness.

\subsubsection{Quantifying the implementability gap} \label{s:methods.opt.gap}

A key metric introduced in this work is the implementability gap, defined as the difference in total energy savings between Strategy~B (theoretical optimum) and Strategy~C (implementable solution):
\begin{equation} \label{eq:impl_gap}
	\Delta_{\mathrm{gap}} = \frac{E_{\mathrm{total}}^C - E_{\mathrm{total}}^B}{E_{\mathrm{total}}^{\mathrm{base}}} \times 100\%
\end{equation}

The corresponding recovery ratio can be calculated as:
\begin{equation} \label{eq:recovery}
	\eta_{\mathrm{recovery}} = \frac{E_{\mathrm{total}}^{\mathrm{base}} - E_{\mathrm{total}}^C}{E_{\mathrm{total}}^{\mathrm{base}} - E_{\mathrm{total}}^B} \times 100\%
\end{equation}
where $E_{\mathrm{total}}^{\mathrm{base}}$, $E_{\mathrm{total}}^B$, and $E_{\mathrm{total}}^C$ are the annual total energy consumption under the baseline, Strategy~B, and Strategy~C, respectively. A high recovery ratio ($\eta_{\mathrm{recovery}} \to 100\%$) indicates that the ramp constraints impose minimal energy penalty relative to the unconstrained optimum, suggesting that the theoretical savings are largely realizable in practice.

\section{Results} \label{s:results}

This section presents the optimization results using a full calendar year (2023) of operational data from the Frontier supercomputer facility at Oak Ridge National Laboratory. The dataset comprises 47{,}186 validated operating points recorded at 10-minute intervals, after filtering for sensor anomalies and maintenance shutdowns.

\subsection{Digital twin validation} \label{s:results.validation}

The digital twin is validated against the full annual dataset ($n = 47{,}186$ records) using the ASHRAE Guideline~14 protocol described in Section~\ref{s:methods.validation}. The predicted return temperature in each subloop follows the energy balance Equation~(\ref{eq:energy_balance}) with $c_p = 3{,}500$~J/(kg$\cdot$K), intentionally distinct from the dataset's internal constant ($c_p = 3{,}709$~J/(kg$\cdot$K)) to ensure a non-circular validation. Table~\ref{tab:validation} reports the per-subloop errors. All three subloops satisfy the stricter targets adopted in this study (CV-RMSE $\leq$ 5\%, $|$NMBE$|$ $\leq$ 5\%). Subloop~3 exhibits the largest CV-RMSE (2.67\%) due to its higher mean temperature differential ($\Delta T_3 = 13.8$~$^\circ$C vs.\ $\Delta T_{1,2} \approx 8$~$^\circ$C), which amplifies sensitivity to the assumed $c_p$. The positive NMBE across all subloops indicates a consistent over-prediction bias of 0.5 - 0.9~$^\circ$C, attributable to the 6\% difference between the adopted and dataset values.

\begin{table}[htbp]
  \centering
  \caption{Digital twin validation: per-subloop return temperature
    prediction errors evaluated over the full annual dataset
    ($n = 47{,}186$; $p = 1$ calibration parameter).}
  \label{tab:validation}
  \begin{tabular}{l c c c}
    \toprule
    \textbf{Metric} & \textbf{Subloop 1} & \textbf{Subloop 2} & \textbf{Subloop 3} \\
    \midrule
    CV-RMSE (\%)              & 1.96    & 1.97    & 2.67    \\
    NMBE (\%)                 & $+$1.69 & $+$1.72 & $+$2.43 \\
    RMSE ($^{\circ}$C)        & 0.551   & 0.558   & 0.907   \\
    $R^2$                     & 0.9933  & 0.9927  & 0.9088  \\
    \midrule
    \multicolumn{4}{l}{\footnotesize ASHRAE Guideline~14 thresholds:
      CV-RMSE $\leq 30\%$, $|\mathrm{NMBE}| \leq 10\%$.} \\
    \multicolumn{4}{l}{\footnotesize Stricter targets adopted here:
      CV-RMSE $\leq 5\%$, $|\mathrm{NMBE}| \leq 5\%$.} \\
    \bottomrule
  \end{tabular}
\end{table}

\subsection{Baseline operating conditions} \label{s:results.baseline}

Figure~\ref{fig:annual_conditions} presents the annual operating profile of the Frontier HTW cooling loop across 2023. The computational heat load remains relatively stable between 7 and 12~MW year-round, reflecting the sustained utilization typical of leadership-class supercomputers. The supply temperature tracks the ambient wet-bulb temperature seasonally (12$^\circ$C in winter to 29$^\circ$C in summer), and the return temperature remains well below the 42$^\circ$C optimization constraint at all times. Most notably, the pump flow rate exhibits a pronounced step change around May, jumping from approximately 200~kg/s to 350 - 400~kg/s and remaining elevated through October. This step-like behavior suggests the pumps are operated at fixed-speed setpoints without continuous optimization.

\subsection{Over-pumping diagnosis} \label{s:results.overpumping}

To quantify excess pumping, the measured baseline flow is compared against the analytical minimum flow required to satisfy $T_\mathrm{return} \leq 42$~$^\circ$C at each CDU branch:
\begin{equation}
\dot{m}_\mathrm{min} = \max_{j \in \{1,2,3\}} \frac{Q_j}{f_j \, c_p \left(T_\mathrm{limit} - T_\mathrm{supply}\right)}
\label{eq:min_flow}
\end{equation}
Figure~\ref{fig:overpumping} confirms systematic over-pumping across all 47{,}186 operating points, with a median over-pumping ratio of 1.5$\times$ (i.e., the facility pumps 50\% more coolant than thermally necessary). Under the cubic affinity law, this 1.5$\times$ excess flow corresponds to a $1.5^3 = 3.4\times$ overspend in pump power.

\subsection{Annual energy comparison across strategies} \label{s:results.annual}

\begin{figure}[h]
	\centering
	\includegraphics[width=\linewidth]{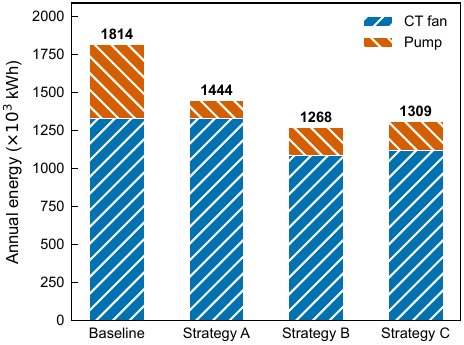}
	\caption{Annual cooling energy breakdown by component (pump vs.\ CT fan) for each optimization strategy. Stacked bars show the relative contributions; total values are annotated in thousands of kWh.}
	\label{fig:energy_breakdown}
\end{figure}

Table~\ref{tab:annual_energy} summarizes the annual energy results for all four
strategies and Figure~\ref{fig:energy_breakdown} provides a visual comparison. A
central finding is that cooling tower (CT) fan energy dominates the total cooling
energy budget, accounting for 73\% of baseline consumption (1{,}325~MWh out of
1{,}814~MWh). Strategy~A optimizes only pump flow, achieving a 75.7\% reduction
in pump energy but leaving the CT component untouched (20.4\% total savings).
Strategy~B introduces supply temperature setpoint reset, raising
$T_\mathrm{supply}$ by an average of 2.7$^\circ$C, which reduces CT fan energy
by 17.9\% and achieves 30.1\% total savings. Strategy~C enforces ramp-rate
constraints and achieves 27.8\% total savings, recovering 92.4\% of the
unconstrained theoretical bound.

\subsection{Monthly energy profiles and supply temperature mechanism} \label{s:results.monthly}

\begin{figure*}[t]
	\centering
	\includegraphics[width=\textwidth]{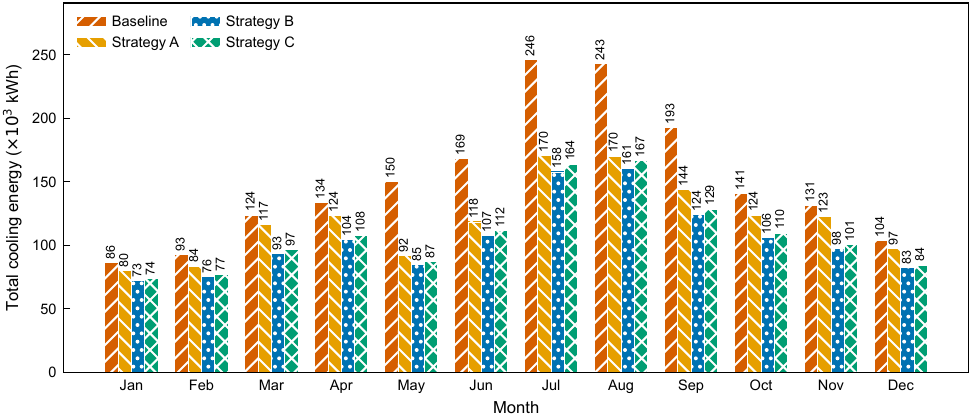}
	\caption{Monthly total cooling energy (pump + CT fan) across four strategies. Strategy~C achieves the largest savings during high-load summer months, with a peak of 33\% reduction in July.}
	\label{fig:monthly_energy}
\end{figure*}

Figure~\ref{fig:monthly_energy} presents the monthly total cooling energy for all four strategies. Winter months (January through March) exhibit savings of 15 - 20\%, constrained by already-low supply temperatures that limit setpoint adjustment. Summer months (May through September) realize the largest savings, with a peak of 33\% in July. The monthly savings heatmap (Figure~\ref{fig:monthly_heatmap} in Appendix~A) confirms that co-optimization strategies outperform flow-only optimization across all months, with the B-to-C gap smallest in summer (1 - 3\%).

Table~\ref{tab:seasonal} quantifies the seasonal breakdown. Summer (June through August) accounts for 43\% of total annual savings under Strategy~C despite spanning only 25\% of the calendar year. The implementability cost varies from 1.5\% in winter to 2.6\% in summer; the relative cost is largest in winter, where smaller absolute savings amplify the impact of constraint-induced lag.

\begin{figure*}[t]
	\centering
	\includegraphics[width=\textwidth]{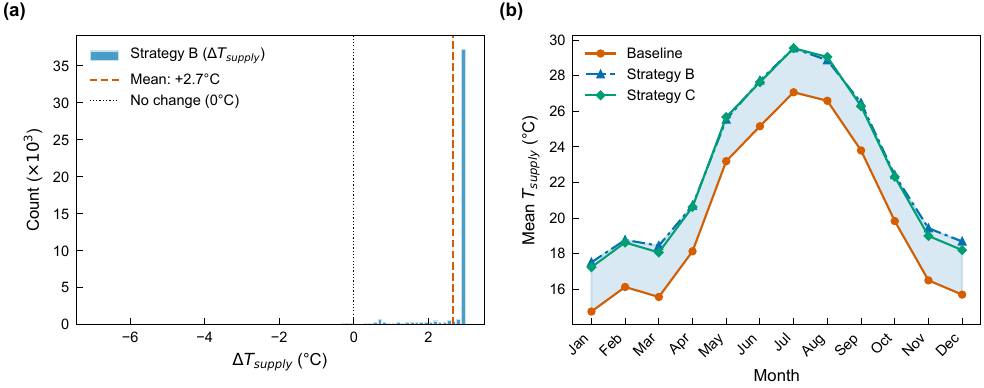}
\caption{Supply temperature setpoint reset mechanism: (a) distribution of $\Delta T_\mathrm{supply}$ showing a mean increase of $+2.7^\circ$C, and (b) monthly mean $T_\mathrm{supply}$ comparing Baseline, Strategy B, and Strategy C. The shaded region highlights the additional headroom exploited by optimization.}\label{fig:tsupply_mechanism}
\end{figure*}

The mechanism behind co-optimization savings is illustrated in Figure~\ref{fig:tsupply_mechanism}. The optimizer raises $T_\mathrm{supply}$ by a mean of $+2.7$~$^\circ$C in 97.9\% of timesteps, with the gap widening in summer when the pump-versus-CT trade-off is most consequential. During winter, the baseline supply temperature is already near the cooling water temperature, leaving limited room for adjustment.

\begin{table*}[t]
	\centering
	\caption{Annual cooling energy comparison across optimization strategies (pump exponent $n=3$). All strategies evaluated using the same optimization framework for consistency.}
	\label{tab:annual_energy}
	\begin{tabular}{l r r r r}
		\toprule
		& \textbf{Baseline} & \textbf{Strategy A} & \textbf{Strategy B} & \textbf{Strategy C} \\

		\midrule
		Pump energy (kWh)      & 488{,}857  & 118{,}772  & 180{,}174  & 193{,}033 \\
		CT fan energy (kWh)    & 1{,}324{,}797 & 1{,}324{,}797 & 1{,}087{,}613 & 1{,}116{,}024 \\
		\textbf{Total (kWh)}   & \textbf{1{,}813{,}654} & \textbf{1{,}443{,}569} & \textbf{1{,}267{,}787} & \textbf{1{,}309{,}057} \\
		\midrule
		Pump savings (\%)      & ---       & 75.7      & 63.1      & 60.5 \\
		CT fan savings (\%)    & ---       & 0.0       & 17.9      & 15.8 \\
		\textbf{Total savings (\%)} & \textbf{---} & \textbf{20.4} & \textbf{30.1} & \textbf{27.8} \\
		\midrule
		Mean flow (kg/s)       & 257       & 163       & 187       & 189 \\
		Mean $T_\mathrm{supply}$ ($^\circ$C) & 20.1 & 20.1 & 22.8 & 22.6 \\
		Max $T_\mathrm{return}$ ($^\circ$C) & ---  & 42.3  & 41.9  & 42.9$^\dagger$ \\
		Thermal violations$^{\dagger\dagger}$ & ---  & 2  & 0  & 9 \\
		\bottomrule
		\multicolumn{5}{l}{\footnotesize $^\dagger$All exceedances occur at the maximum pump capacity (420~kg/s),}\\
		\multicolumn{5}{l}{\footnotesize indicating a system capacity limit rather than an optimization failure.}\\
		\multicolumn{5}{l}{\footnotesize $^{\dagger\dagger}$Points exceeding the 42$^\circ$C optimization constraint; all remain below the 45$^\circ$C equipment limit.}
	\end{tabular}
\end{table*}

\begin{table*}[t]
	\centering
	\caption{Seasonal cooling energy comparison and total savings
	  by strategy.}
	\label{tab:seasonal}
	\begin{tabular}{l r r r r r r r}
		\toprule
		& \multicolumn{4}{c}{\textbf{Energy (MWh)}}
		& \multicolumn{3}{c}{\textbf{Savings (\%)}} \\
		\cmidrule(lr){2-5} \cmidrule(lr){6-8}
		& \textbf{Baseline} & \textbf{A} & \textbf{B} & \textbf{C}
		& \textbf{A} & \textbf{B} & \textbf{C} \\
		\midrule
		Winter (Dec - Feb)  & 280 & 259 & 229 & 233
		  &  7.6 & 18.5 & 17.0 \\
		Spring (Mar - May)  & 407 & 333 & 284 & 293
		  & 18.1 & 30.3 & 28.0 \\
		Summer (Jun - Aug)  & 659 & 458 & 425 & 442
		  & 30.5 & 35.5 & 33.0 \\
		Fall (Sep - Nov)    & 467 & 393 & 331 & 341
		  & 15.9 & 29.2 & 26.9 \\
		\midrule
		\textbf{Annual}    & \textbf{1{,}814} & \textbf{1{,}444}
		  & \textbf{1{,}268} & \textbf{1{,}309}
		  & \textbf{20.4} & \textbf{30.1} & \textbf{27.8} \\
		\bottomrule
	\end{tabular}
\end{table*}

\subsection{Operating envelope shift} \label{s:results.envelope}

Co-optimization shifts the operating envelope from the baseline centroid (257~kg/s, 20.1$^\circ$C) toward lower flow and higher supply temperature: Strategy~B centers at 187~kg/s and 22.8$^\circ$C, while Strategy~C centers at 189~kg/s and 22.6$^\circ$C. The joint distribution plots (Figure~\ref{fig:joint_distribution} in Appendix~A) confirm that both strategies eliminate the baseline bimodal flow structure and compress the operating envelope. The near-identical centroids of Strategies~B and~C indicate that the ramp constraints primarily affect transition dynamics rather than the steady-state operating target.

\subsection{Strategy~C dynamic behavior under constraints} \label{s:results.dynamic}

\begin{figure*}[t]
	\centering
	\includegraphics[width=\textwidth]{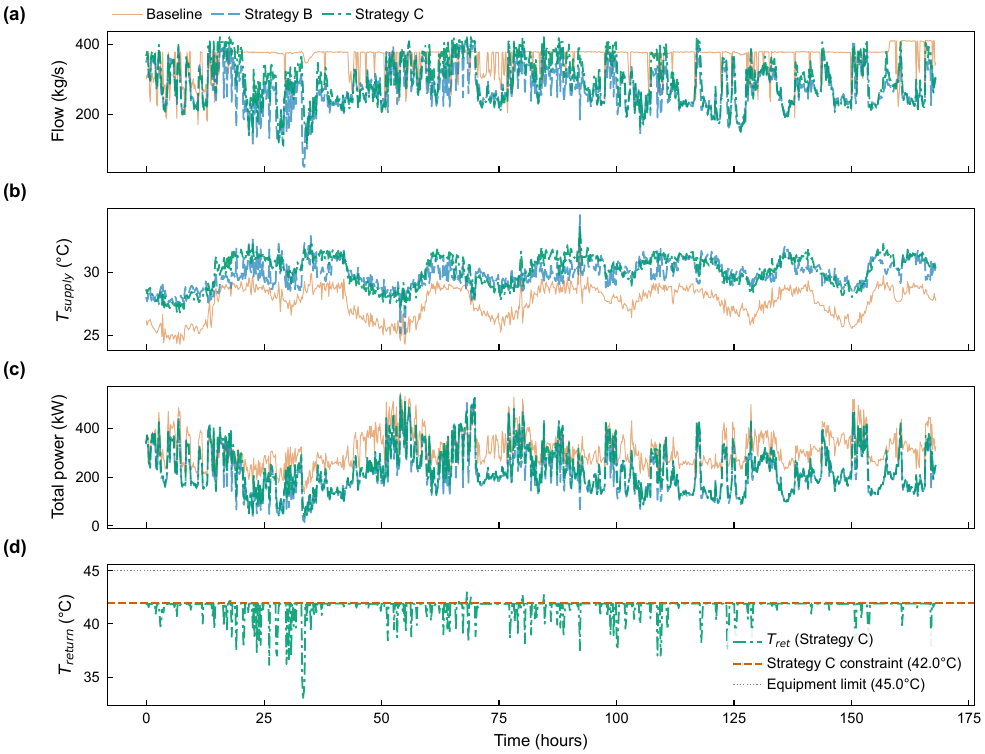}
	\caption{Representative summer week (July) comparing Baseline, Strategy~B, and Strategy~C: (a)~flow rate, (b)~supply temperature setpoint, (c)~total cooling power, and (d)~return temperature with constraint thresholds. Strategy~C maintains near-complete constraint compliance while achieving significant power reduction.}
	\label{fig:summer_week}
\end{figure*}

Figure~\ref{fig:summer_week} presents a representative summer week (168 hours in July), comparing the dynamic trajectories of Baseline, Strategy~B, and Strategy~C. Strategy~C produces smoother control trajectories than Strategy~B while maintaining a consistently higher supply temperature setpoint (28 - 32$^\circ$C vs.\ baseline 24 - 28$^\circ$C), achieving 30 - 50\% lower instantaneous total power during most of the week. The return temperature operates near the 42$^\circ$C constraint throughout, with the few exceedances occurring exclusively at maximum pump capacity (420~kg/s) where the system has exhausted its flow control authority. All operating points remain below the 45$^\circ$C equipment limit.

\subsection{Ramp-rate compliance and pump power distribution} \label{s:results.ramp}

A practical concern for real-world deployment is whether optimization commands can be tracked by physical actuators. Strategy~B produces flow rate changes up to 189~kg/s per step and temperature changes up to 4.4$^\circ$C per step, frequently violating the $\pm 50$~kg/s and $\pm 1$~$^\circ$C limits that represent conservative actuator constraints informed by variable-frequency drive slew rates \citep{ASHRAE2020}. Strategy~C respects both limits by construction (Figure~\ref{fig:ramp_compliance} in Appendix~A), achieving 27.8\% total savings versus 30.1\% for Strategy~B. The annual pump power duration curves (Figure~\ref{fig:duration_curve} in Appendix~A) confirm that all optimized strategies substantially reduce pump power relative to the baseline, with Strategy~A achieving the largest pump reduction and Strategies~B and~C trading slightly higher pump power for cooling-tower fan savings.

\subsection{Sensitivity to pump curve exponent} \label{s:results.sensitivity}

Because the exact Frontier pump curves are not publicly available, all strategies are evaluated across three representative exponents ($n \in \{2.0, 2.5, 3.0\}$). Table~\ref{tab:sensitivity} presents the results, while Figure~\ref{fig:sensitivity} in Appendix~A provides a visual comparison. Under the most conservative assumption ($n = 2$), Strategy~C achieves 21.3\% total savings; under the standard affinity law ($n = 3$), savings reach 27.8\%. This 21 - 28\% range provides a robust bound on expected benefits. The relative advantage of co-optimization over flow-only optimization is most pronounced at lower pump exponents: at $n = 2$, Strategy~C nearly doubles the savings of Strategy~A (21.3\% vs.\ 10.9\%), because cooling-tower fan savings from setpoint reset become the dominant contributor when pump power is less sensitive to flow.

\begin{table}[htbp]\small\setlength{\tabcolsep}{4pt}
	\centering
	\caption{Total cooling energy savings (\%) across pump curve exponents and strategies.}
	\label{tab:sensitivity}
	\begin{tabular}{l c c c}
		\toprule
		\textbf{Pump exponent} & \textbf{Strategy A} & \textbf{Strategy B} & \textbf{Strategy C} \\
		\midrule
		$P \propto \dot{m}^2$ (conservative) & 10.9 & 23.3 & 21.3 \\
		$P \propto \dot{m}^{2.5}$ (typical)  & 15.2 & 26.3 & 24.2 \\
		$P \propto \dot{m}^3$ (affinity law)  & 20.4 & 30.1 & 27.8 \\
		\bottomrule
	\end{tabular}
\end{table}

\section{Discussion} \label{s:discussion}

\subsection{System-level perspective on cooling system optimization}
\label{s:discussion.system}

A primary contribution of this work is the demonstration that optimizing pump flow alone captures only a fraction of the available cooling energy savings. The finding that CT fan energy constitutes 73\% of the total cooling energy budget fundamentally shifts the optimization landscape. While the pump affinity law ($P \propto \dot{m}^3$) makes flow reduction an attractive target in isolation, the system-level optimum lies in coordinating pump flow and supply temperature setpoint to minimize the combined pump and CT fan consumption. This coordination produces a counterintuitive outcome: the co-optimized Strategy~B uses more pump energy than the flow-only Strategy~A (180,174 vs.\ 118,772~kWh), yet achieves greater total savings (30.1\% vs.\ 20.4\%) by reducing the dominant CT fan component. This finding aligns with the broader principle that component-level optimization can be suboptimal when interdependencies exist between subsystems, a pattern likely to recur in other coupled thermal systems where multiple energy-consuming components share hydraulic and thermal coupling.

\subsection{Practical implementability of the ramp-constrained strategy}
\label{s:discussion.impl}

Strategy~C addresses a gap often observed in optimization studies that report unconstrained theoretical savings without considering actuator limitations. The $\pm 50$~kg/s flow rate limit and $\pm 1$~$^\circ$C setpoint change limit per 10-minute step represent conservative bounds informed by typical variable-frequency drive slew rates and thermal inertia considerations for large-scale cooling systems \citep{ASHRAE2020}. Strategy~C recovers 92.4\% of the unconstrained savings (27.8\% vs.\ 30.1\%) while producing smooth, actuator-friendly control trajectories. The 2.3\% reduction from the theoretical bound is a modest penalty for a strategy deployable on real hardware without risking water hammer, thermal shock, or control oscillations. The seasonal analysis (Table~\ref{tab:seasonal}) provides additional nuance: the implementability gap ranges from 1.5\% in winter to 2.6\% in summer, with the relative impact proportionally larger during winter when limited $T_\mathrm{supply}$ headroom slows convergence to the optimal setpoint. A seasonally adaptive ramp-rate policy could further narrow this gap.

\subsection{Robustness and generalizability}
\label{s:discussion.robustness}

The 21 - 28\% savings range across pump exponents (Table~\ref{tab:sensitivity}) provides confidence that the proposed approach is robust to uncertainty in the pump power model, which is practically significant because detailed pump curves for installed equipment are often unavailable. Even under the most conservative assumption ($P \propto \dot{m}^2$), Strategy~C achieves over 21\% total cooling energy savings (approximately 380~MWh annually for the Frontier facility). The methodology is applicable to other liquid-cooled data centers that employ external heat exchangers and cooling towers, provided that sufficient operational data are available for digital twin calibration. The Modelica-based framework enables component substitution (e.g., replacing counterflow with crossflow heat exchangers) without modifying the optimization algorithm, and the $\varepsilon$-NTU analytical model solves each 10-minute optimization in under 10~ms, making the approach suitable for online supervisory control.

\subsection{Limitations and future work} \label{s:discussion.limitations}

Several limitations should be noted. The cooling tower fan model employs a simplified power-approach relationship rather than a manufacturer-specific model, so CT fan savings should be interpreted as directionally correct rather than exact. The optimization assumes that the supply temperature setpoint can be independently adjusted from cooling tower operation, which may require control system modifications. The current study uses static point-by-point optimization with retrospective ramp-rate enforcement; a full receding-horizon MPC formulation with multi-step look-ahead could further improve performance during rapid load transients. The digital twin validation was performed using a steady-state energy balance with an adopted $c_p$ that differs by 6\% from the dataset's internal conversion constant; transient validation during load ramps and independent calorimetric measurement of $c_p$ would strengthen confidence in the absolute predictions.

Future work will focus on three extensions: (1)~integration of weather forecasting to enable predictive setpoint scheduling, (2)~experimental validation through a pilot deployment on a subset of CDU branches, and (3)~extension of the framework to facilities with chilled-water systems, where the optimization trade-offs between chiller and cooling tower energy present analogous coupling effects.

\section{Conclusion} \label{s:Conclusion}

This paper developed a Modelica-based digital twin for the Frontier exascale supercomputer's liquid cooling system and evaluated three progressively constrained optimization strategies across one full year of operational data. The study produced four principal findings.

First, cooling tower fan energy accounts for 73\% of the total cooling energy budget, making supply temperature co-optimization essential. Flow-only optimization (Strategy~A) achieves 20.4\% total savings, but co-optimization of flow and supply temperature (Strategy~B) reaches 30.1\% through a mean supply temperature increase of 2.7$^\circ$C. The co-optimized strategy uses more pump energy yet achieves greater total savings, underscoring the importance of system-level optimization over component-level approaches.

Second, the implementability gap is surprisingly small. Strategy~C, enforcing ramp-rate constraints of $\pm 50$~kg/s per step and $\pm 1$~$^\circ$C per step, recovers 92.4\% of the unconstrained savings (27.8\% vs.\ 30.1\%). This demonstrates that near-optimal energy performance is achievable without sacrificing actuator smoothness.

Third, the methodology is robust to parametric uncertainty: across pump power exponents $n = 2$ to $n = 3$, Strategy~C achieves 21 - 28\% total savings, providing a reliable bound for facilities lacking detailed pump curve data.

Fourth, a physics-based digital twin with a single calibration parameter (glycol mass fraction) predicts per-subloop return temperatures to within 1.96 - 2.67\% CV-RMSE, confirming the viability of equation-based digital twins as optimization testbeds for HPC cooling infrastructure.

The implementability gap metric and layered optimization framework introduced here are applicable to any coupled thermal system where the cost of practical constraints on theoretically optimal control must be quantified. As HPC facilities scale toward multi-exascale power envelopes, systematic measurement of this gap will become increasingly important.

\section*{Acknowledgment}
This research is partially supported from the University of Michigan-Dearborn Office of Research through Research Initiation \& Development (RID) Grant.

\section*{Data and code availability}
The data and computer code used in this study are not yet publicly available, but will be made publicly available at the time of paper acceptance.

\bibliographystyle{unsrtnat}
\bibliography{ref}

\clearpage
\appendix
\renewcommand{\thefigure}{A\arabic{figure}}
\setcounter{figure}{0}
\renewcommand{\thetable}{A\arabic{table}}
\setcounter{table}{0}

\section{Supplementary Material}
\label{app:supplementary}

\section*{Appendix A}
\addcontentsline{toc}{section}{Appendix A}

\subsection{Data center cooling system optimization: approaches and limitations} \label{s:intro.optimization}

\begin{table*}[t]
	\centering
	\caption{Comparison of representative cooling system optimization studies.
	  Abbreviations: DT = digital twin; HPC = high-performance computing; DC = data center.}
	\label{tab:lit_comparison}
	\small
	\begin{tabular}{@{}l l l c c c@{}}
		\toprule
		\textbf{Study} & \textbf{Method} & \textbf{Application}
		  & \textbf{DT} & \textbf{Constraints}
		  & \textbf{Gap metric} \\
		\midrule
		Evans and Gao~\citep{Evans2016}   & Neural net       & Google DC
		  & No       & No      & No  \\
		Lazic et al.~\citep{Lazic2018}   & MPC              & Google DC
		  & No       & Partial & No  \\
		Li et al.~\citep{Li2020}      & Deep RL          & Enterprise DC
		  & No       & No      & No  \\
		Sarkar et al.~\citep{Sarkar2024}  & Multi-agent RL   & Enterprise DC
		  & Yes      & No      & No  \\
		Zhan et al.~\citep{Zhan2025}    & Offline RL       & Production DC
		  & No       & Partial & No  \\
		Zhu and Lin~\citep{Zhu2024}     & Real-time opt.   & Enterprise DC
		  & Yes      & Partial & No  \\
		Jadhav and Liu~\citep{Jadhav2026}  & ML surrogate     & Frontier (HPC)
		  & No       & No      & No  \\
		Brewer et al.~\citep{Brewer2024}  & --- (V\&V only)  & Frontier (HPC)
		  & Modelica & ---     & No  \\
		Fu et al.~\citep{Fu2019a}     & --- (Sim.\ only) & Air-cooled DC
		  & Modelica & ---     & No  \\
		\textbf{This study} & \textbf{SLSQP (layered)}
		  & \textbf{Frontier (HPC)}
		  & \textbf{Modelica} & \textbf{Ramp-rate}
		  & \textbf{Yes} \\
		\bottomrule
	\end{tabular}
\end{table*}

\subsection{Monthly savings heatmap by strategy} \label{app:supplementary}

\begin{figure*}[t]
	\centering
	\includegraphics[width=\textwidth]{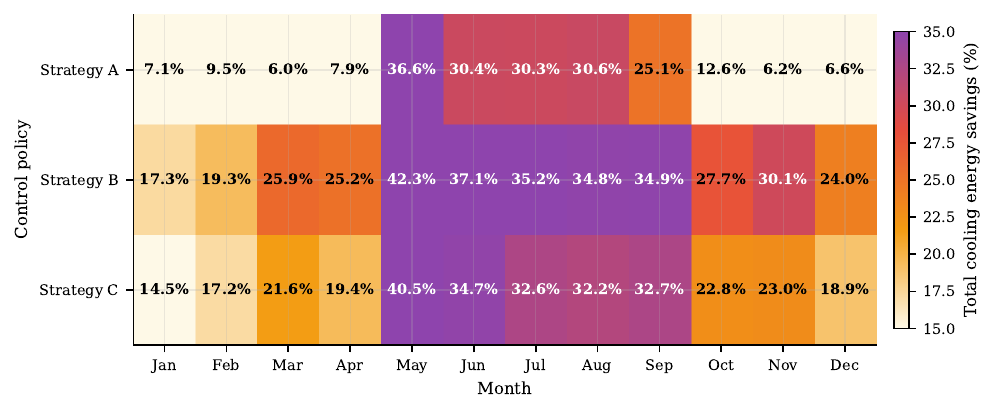}
	\caption{Monthly total cooling energy savings (\%) by control policy. Color intensity indicates magnitude. Co-optimization strategies (B and C) outperform flow-only optimization (A) across all months, with the advantage most pronounced during shoulder and winter seasons.}
	\label{fig:monthly_heatmap}
\end{figure*}

Figure~\ref{fig:monthly_heatmap} provides a compact month-by-month view of cooling-energy savings across all optimization strategies. The heatmap makes the seasonal structure more explicit than the monthly bar chart in the main text. Strategy~A contributes relatively modest savings during winter (approximately 6 - 10\%), when the baseline operating point is already close to thermally efficient conditions and the available margin for flow reduction is limited. Its contribution rises substantially during summer, reaching about 25 - 37\%, when baseline flow is most inflated relative to thermal need.

Strategies~B and~C consistently outperform Strategy~A across the year because co-optimization exploits both flow reduction and supply-temperature reset. Their largest absolute advantage over Strategy~A appears during the shoulder seasons (March - April and October - November), where setpoint adjustment remains effective even when incremental pump savings are smaller than in summer. The difference between Strategies~B and~C is smallest during summer, typically around 1 - 3\%, indicating that the ramp-rate constraints impose the least penalty precisely when the achievable savings are greatest.

Overall, the appendix heatmap complements Figure~\ref{fig:monthly_energy} by showing that the annual savings reported in the main text are not driven by a few isolated months, but instead reflect a consistent seasonal pattern in which co-optimization provides benefits throughout the year while remaining most effective during high-load warm-weather operation.

\subsection{Operating envelope shift} \label{app:envelope_shift}

\begin{figure*}[t]
	\centering
	\includegraphics[width=\textwidth]{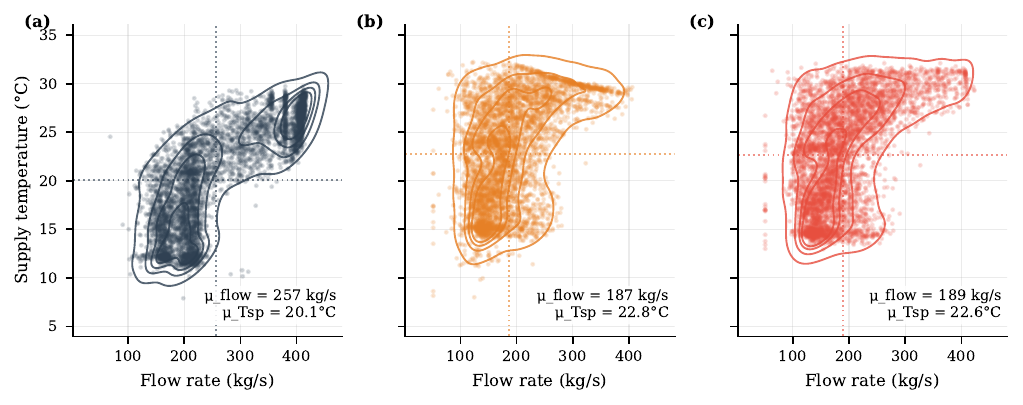}
	\caption{Joint distribution of flow rate and supply temperature for (a)~Baseline, (b)~Strategy~B, and (c)~Strategy~C. Contour lines show kernel density estimates; dotted lines indicate mean values. Co-optimization shifts the operating envelope toward lower flow and higher supply temperature, while Strategy~C closely tracks the Strategy~B centroid with modestly broader spread due to ramp-rate constraints.}
	\label{fig:joint_distribution}
\end{figure*}

Figure~\ref{fig:joint_distribution} presents the joint distribution of flow rate and supply temperature for Baseline, Strategy~B, and Strategy~C, visualized through kernel density contours overlaid on the operating points. Under baseline operation, the facility occupies a broader envelope centered at 257~kg/s mean flow and 20.1$^\circ$C mean supply temperature, with a bimodal flow structure consistent with the seasonal step change discussed in the main text.

Strategy~B shifts the operating envelope toward lower flow rates and higher supply temperatures, centering at 187~kg/s and 22.8$^\circ$C. Strategy~C exhibits a nearly identical centroid at 189~kg/s and 22.6$^\circ$C, but with a modestly broader spread, reflecting the smoothing effect of ramp-rate constraints. The close similarity between Strategies~B and~C confirms that ramp constraints primarily affect transition dynamics rather than the steady-state operating target.

\subsection{Ramp-rate compliance} \label{app:ramp_compliance}

\begin{figure*}[t]
	\centering
	\includegraphics[width=\textwidth]{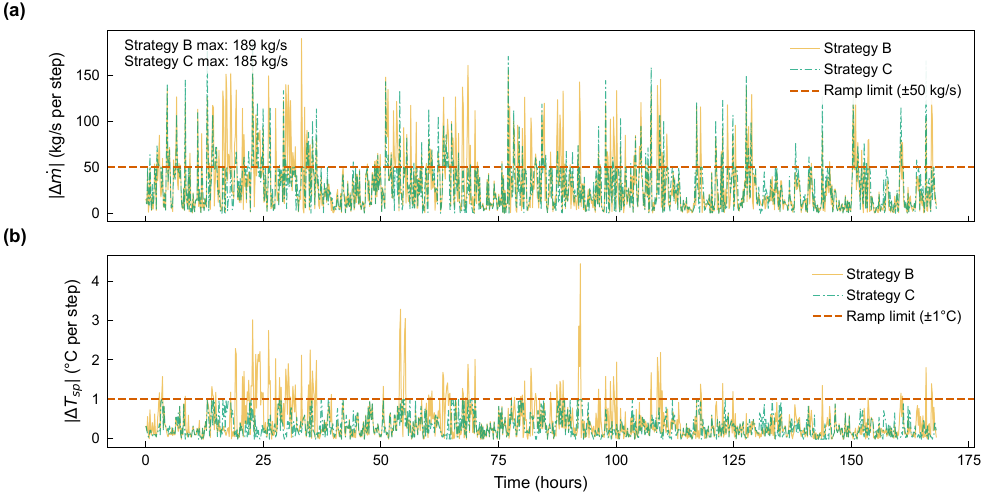}
	\caption{Step-to-step ramp-rate comparison between unconstrained Strategy~B and ramp-constrained Strategy~C during a summer week: (a)~flow rate changes with $\pm$50~kg/s limit, and (b)~$T_\mathrm{supply}$ setpoint changes with $\pm$1$^\circ$C limit. Strategy~C respects both operational constraints while Strategy~B frequently violates them.}
	\label{fig:ramp_compliance}
\end{figure*}

Figure~\ref{fig:ramp_compliance} evaluates whether the optimization commands remain compatible with conservative actuator and process limits. The unconstrained Strategy~B produces large step-to-step flow and supply-temperature changes that frequently exceed the adopted limits of $\pm 50$~kg/s and $\pm 1$~$^\circ$C per 10-minute interval. By contrast, Strategy~C satisfies both limits by construction, indicating that the ramp-constrained policy remains practically implementable while retaining most of the achievable energy benefit.

\subsection{Pump power duration curves} \label{app:duration_curve}

\begin{figure*}[t]
	\centering
	\includegraphics[width=0.85\textwidth]{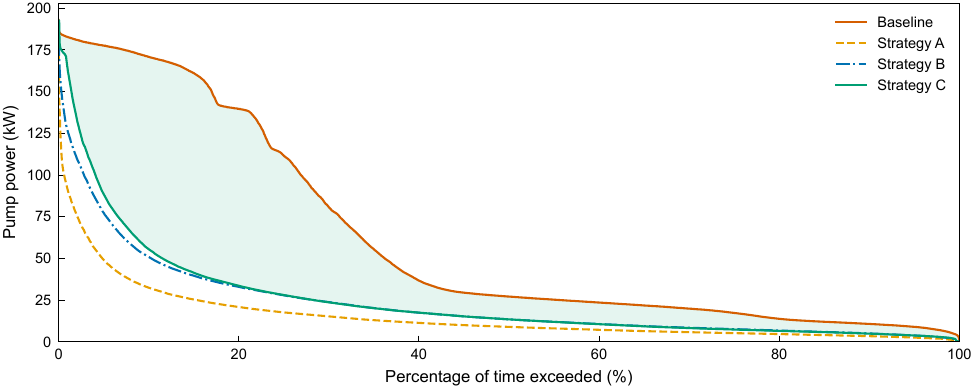}
	\caption{Annual pump power duration curves showing the distribution of pump power across all operating hours. All optimization strategies substantially reduce pump power relative to the baseline, with the greatest reductions during high-load summer operation.}
	\label{fig:duration_curve}
\end{figure*}

Figure~\ref{fig:duration_curve} shows the annual pump power duration curves for the baseline and all optimized strategies. All optimization strategies shift the curve downward relative to the baseline, confirming substantial reduction in pump power across operating hours. The largest reduction is achieved by Strategy~A, whereas Strategies~B and~C retain slightly higher pump power because they trade part of the pump savings for lower cooling-tower fan energy.

\subsection{Sensitivity to pump curve exponent} \label{app:sensitivity}

\begin{figure*}[t]
	\centering
	\includegraphics[width=0.85\textwidth]{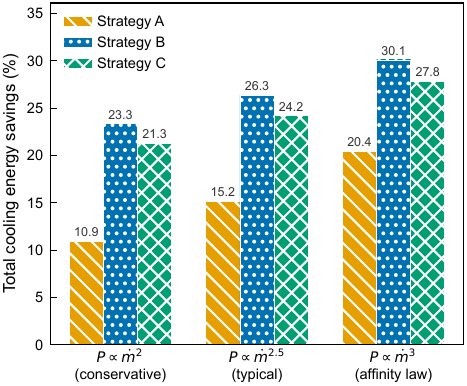}
	\caption{Sensitivity of total cooling energy savings to pump curve exponent for all three optimization strategies. Co-optimization (Strategies~B and~C) consistently outperforms flow-only optimization across all exponents.}
	\label{fig:sensitivity}
\end{figure*}

Figure~\ref{fig:sensitivity} shows the sensitivity of total cooling-energy savings to the assumed pump curve exponent for all three optimization strategies. Savings increase monotonically with exponent across all strategies, consistent with the nonlinear dependence of pump power on flow rate. Co-optimization strategies (B and C) consistently outperform flow-only optimization (A) across the full range of plausible exponents, with the relative advantage most pronounced at lower exponents where cooling-tower fan savings account for a larger share of the total benefit.

\end{document}